\documentclass[9pt,twocolumn,twoside]{osajnl}

\journal{ol} 

\setboolean{shortarticle}{true}

\title{Sum frequency generation from touching wires: A transformation optics approach}

\author[1,*]{Shimon Elkabetz}
\author[1]{K. Nireekshan Reddy}
\author[1]{Y. Sivan}

\affil[1]{School of Electrical and Computer Engineering, Ben-Gurion, Be'er-Sheva 8410501, Israel}

\affil[*]{Corresponding author: elshimon@post.bgu.ac.il}

\ociscodes{(190.0190) Nonlinear optics, (190.4350) Nonlinear optics at surfaces, (190.4223)  Nonlinear wave mixing, (240.6680) Surface plasmons.}

\begin{abstract}
We employ transformation optics to study analytically nonlinear wave mixing from a singular geometry of touching plasmonic wires. We obtain the analytic solution of the near-field and complement it with a solution of the far-field properties. We find, somewhat surprisingly, that optimal efficiency (in both regimes) is obtained for the degenerate case of the Second-harmonic generation. We exploit the analytic solution obtained to trace this behaviour to the spatial overlap of the input fields near the geometric singularity. 
\end{abstract}

\setboolean{displaycopyright}{true}

\begin{document}

\maketitle

\section{Introduction}

Second-order nonlinear optical wave interactions involve the coherent conversion of two input optical waves into a third wave whose frequency is the sum of that of the input waves~\cite{Boyd-book}. This enables frequency conversion of optical waves, a phenomenon of both fundamental and practical importance~\cite{Boyd-book}.

As the strength of second-order optical nonlinear interactions is weak in most materials, efforts have been dedicated to finding ways to increase it. One of the promising such ways is the use of metal (plasmonic) nanostructures. The high local-fields enabled by these structures can be of particular benefit for this purpose. Since most plasmonic materials are centrosymmetric, second-order nonlinear optical processes in metals are forbidden in the local bulk response~\cite{Boyd-book}, and thus are governed by surface symmetry-breaking effects~\cite{Bloembergen:1968vc,Rudnick:1971wb,Sipe:1980vz}. 

Analytic solutions for the field distribution near plasmonic nanostructures are available just for the simplest structures. The complex second-order response makes such solutions even more rare for nonlinear wave interactions. Accordingly, the vast majority of analytic studies of nonlinear wave interactions in plasmonic nanostructures rely on numerical simulations.

\begin{figure}
\centering\includegraphics[width=1 \linewidth]{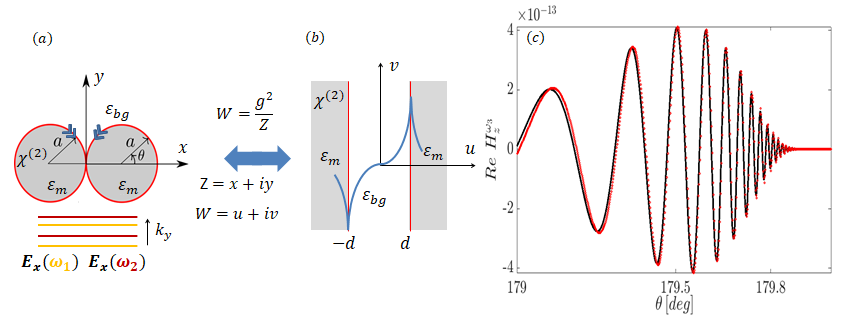}
\caption{(Color online) (a) A touching wire dimer and (b) slab structure related through conformal inversion transformation. The incident plane waves $E^{\omega_{1,2}}_x$ are marked by red and yellow solid lines,~$\chi^{(2)}$ is the second order nonlinear surface tensor which marked by red lines along the structures contours, $\varepsilon_{m/bg}$ are the permittivities of the metal/background, $d=g^2/2a$ is the distance of the interfaces from the origin where $a$ is the cylinder radius and $g^2$ is a scaling constant. (c) Analytic (black solid line) and numeric (red dots) solutions of $H_{z,TW}^{\omega_3}$ on the perimeter of the TWs as a function of the angle $\theta$ (see Fig.~\ref{TWandslab}), $\omega_3$ = $908$ THz and $\omega_1$ = $375$ THz and $\omega_2$ = $533$ THz. }
\label{TWandslab}
\end{figure}

Recently, we studied analytically surface second-harmonic generation (SHG) assisted by surface plasmon polaritons (SPPs) from nano metallic touching wires (TWs) surrounded by a transparent dielectric medium~\cite{Reddy_SHG_TO}. This structure is interesting because it is a singular plasmonic structure, known for its optimal ability to enhance the electromagentic fields close the touching point by several orders of magnitude~\cite{alex_kissing_NJP,Paloma-review-TO}; they also provide unusually wide spectral response.  
The unique analytic solution was enabled by the technique of conformal transformation optics (TO) under which Maxwell's equations are invariant and the material and spectral characteristics of the system are preserved, see Fig.~\ref{TWandslab}. Similar to the linear case, we transform to the simpler slab structure, solve and then transform back to the more complicated TWs structure. This unique approach, one of the first ever to employ TO to nonlinear wave interactions, revealed rich physical insights. Specifically, our analysis demonstrated that apart from the mode-matching condition, the phase-matching condition is relevant even for this subwavelength structure. Furthermore, we identified a geometric factor which was not identified before and causes the field suppression close to the touching point~\cite{Reddy_SHG_TO}. In~\cite{Optimization} we showed how to compute the scattering cross section from the analytic solution obtained in~\cite{Reddy_SHG_TO} and performed an optimization of the SHG efficiency by tuning the background permittivity.
In this manuscript, we attempt to further optimize the frequency conversion process by extending the analysis to (non-degenerate) three wave mixing, in particular, to Sum Frequency Generation (SFG). This provides one additional degree of freedom that is a-priori expected to enable further improvement of the conversion efficiency. To do that, we adapt the procedure presented in~\cite{Reddy_SHG_TO}, and then analyze the final solution.
\section{Configuration of study}
We are interested in the analytic solution for three wave mixing from subwavelength TWs. Consider TM ($x$) polarized plane waves $E^{\omega_{1,2}}_{inc,x}$ at the fundamental frequencies (FF) $\omega_{1,2}$, which are incident perpendicularly upon a subwavelength pair of infinitely-long touching metallic wires, see Fig.~\ref{TWandslab}. 
The Sum Frequency magnetic field $H_z^{\omega_3}$ is obtained by solving the Helmholtz equation under the quasistatic (QS) approximation ($\lim k_0 \rightarrow 0$)~\cite{Reddy_SHG_TO}, namely,
\begin{eqnarray}
\nabla \times \left[ \nabla \times H_{z,TW}^{\omega_3}(x,y) \hat{z} \right] = 0, \label{eq:helmholtz}
\end{eqnarray}
where $x$ and $y$ are the in-plane spatial coordinates. The boundary conditions for the electric and magnetic fields associated with the problem, to be applied on the metal-dielectric interfaces are given by~\cite{Reddy_Sivan_SHG_BCs}
\begin{eqnarray}
\mathbf{n}\times(\mathbf{H}_{bg} - \mathbf{H}_{m}) &=& H_{z,TW}^{\omega_3}|_{bg} - H_{z,TW}^{\omega_3}|_m = 0, \\
E_{\parallel,TW}^{\omega_3}|_{bg} - E_{\parallel,TW}^{\omega_3}|_m &=& - \frac1{\varepsilon_{bg}} \nabla_\parallel P_{S,\perp}^{\omega_3}, \label{EbC} 
\end{eqnarray}
where  $\mathbf{n}$ is the outward normal to the interface and the subscripts $bg$ and $m$ correspond to the dielectric and metal sides of the interface, respectively. The SFG frequency is given by $\omega_3 = \omega_1 + \omega_2$ and $\varepsilon_{bg}$ is the background permittivity (assumed to be dispersionless). $P_{S,\perp}^{\omega_3}$ describes the normal to the surface component of the surface polarization vector, $\mathbf{P}_S^{(2)}$~\cite{Reddy_Sivan_SHG_BCs}.

For Sum-frequency generation, the nonlinear surface tensor $\chi_S^{(2)}$ contains three independent elements, the most dominant element of which is $\chi^{(2)}_{S,\perp\perp\perp}$~\cite{NonlinearChi}. This element describes perpendicular surface currents, therefore $P^{\omega_3}_{S,\perp}$ definition given by
\begin{eqnarray}
P^{\omega_3}_{S,\perp} = \chi^{(2)}_{S,\perp\perp\perp} ~E_{\perp,TW}^{\omega_1}~E_{\perp,TW}^{\omega_2} 
~\delta(x^2 + y^2 \mp 2 a x). \label{eq:polarization}
\end{eqnarray}
where $E_{\perp,TW}^{\omega_{1,2}}$ are the electric fields at the fundamental frequencies, $\omega_{1,2}$, perpendicular to the perimeter of the touching cylinder geometry~\cite{alex_kissing_NJP}. The source can be more conveniently written as a magnetic surface current density~\cite{BEMmathod}, given by
\begin{eqnarray}
J_z = - \frac1{\varepsilon_{bg}} \nabla_\parallel P_{S,\perp}^{\omega_3}, \label{EbCwithJ} 
\end{eqnarray}

The boundary condition for the electric field, Eq.~(\ref{EbC}), in terms of the magnetic field is given by
\begin{eqnarray}
&&\left(\frac{x \pm a}{a}\right)\left[\frac{1}{\varepsilon_{bg}} \frac{\partial H^{\omega_3}_{z,TW}}{\partial x}\Big|_{bg} - \frac{1}{\varepsilon_m^{\omega_3}} \frac{\partial H^{\omega_3}_{z,TW}}{\partial x}\Big|_m \right] + \\ \nonumber &&\frac{y}{a} \left[\frac{1}{\varepsilon_{bg}} \frac{\partial H^{\omega_3}_{z,TW}}{\partial y}\Big|_{bg} - \frac{1}{\varepsilon_m^{\omega_3}} \frac{\partial H^{\omega_3}_{z,TW}}{\partial y}\Big|_m \right]  =  i\frac{\omega_3 \varepsilon_0}{\varepsilon_{bg}}\nabla_\parallel P_{S,\perp}^{\omega_3} ,
\end{eqnarray}
where $\varepsilon_0$ is the vacuum permittivity and $\varepsilon_m^{\omega_3}$ is the SFG metallic permittivity. 

According to~\cite{alex_kissing_NJP}, under the QS approximation, the linear electric response field, $\mathbf{E}^{\omega_{1,2}}$ which appear in Eq.~(\ref{eq:polarization}), can be derived by differentiating the electric potential $\phi^{\omega_{1,2}}$. Specifically,
\begin{eqnarray}
\mathbf{E}^{\omega_{1,2}} = {\nabla_{x,y}}\phi^{\omega_{1,2}},
\end{eqnarray}
\begin{eqnarray}
\phi^{\omega_{1,2}} = - \left(\frac{a \pi\varepsilon_{bg}}{\varepsilon_m + \varepsilon_{bg}}\right) \exp\left(\dfrac{-\alpha^{{\omega_{1,2}}}ax}{x^2 + y^2}\right)\exp\left(\dfrac{i\alpha^{{\omega_{1,2}}}a|y|}{x^2 + y^2}\right),\label{eq:potential}
\end{eqnarray}
where $a$ is the radius of a single wire and  $\varepsilon_m^{\omega_{\tiny{1,2}}}$ are the metal permittivities at FF and $\alpha^{\omega_{1,2}}$ are the corresponding dimensionless propagation constants, given by 
\begin{eqnarray}
\alpha^{\omega_{1,2}} = \ln\left(\frac{\varepsilon_m^{\omega_{\tiny{1,2}}} - \varepsilon_{bg}}{\varepsilon_m^{\omega_{\tiny{1,2}}} + \varepsilon_{bg}}\right) ,\quad  \textrm{Re}[\varepsilon_m^{\omega_{\tiny{1,2}}}] < - \varepsilon_{bg}. \label{eq:alpha}
\end{eqnarray}
We note that the form of Eq.~(\ref{eq:alpha}) is just the dispersion relation of SPPs in an metal-dielectric-metal structure under QS approximation.

Using Eq.~(\ref{eq:potential}), the magnetic surface current in the TW frame~(\ref{eq:polarization}) can be rewritten as
\begin{eqnarray}
J_{z,r/l}(x,y) = \frac{\chi^{(2)}_{S,\perp\perp\perp}}{\varepsilon_{bg}}~\partial_\parallel  (E_{\perp,TW}^{\omega_1}E_{\perp,TW}^{\omega_2})~\delta(x^2 + y^2\mp2ax), \label{magneticcurrents}
\end{eqnarray}
where $J_{z,r}$ and $J_{z,l}$ are the magnetic current densities on the right and left cylinders, respectively. Since the parallel derivative does not change the complex exponential function, the magnetic surface currents can be decomposed as
\begin{eqnarray}
J_{z,r}(x,y) &=& \mathcal{R}(x,y)\exp\left(\frac{i \alpha_s a|y|}{x^2 + y^2}\right)~\delta(x^2 + y^2-2ax), \label{eq:JR} \\
J_{z,l}(x,y) &=& \mathcal{L}(x,y)\exp\left(\frac{i \alpha_s a|y|}{x^2 + y^2}\right)~\delta(x^2 + y^2+2ax). \label{eq:JL}
\end{eqnarray}
The FF electric fields (see Eq.~(\ref{eq:potential}) are symmetric in $x$ and $y$, so that their product is symmetric as well. According to Eq.~(\ref{magneticcurrents}), the parallel derivative changes this symmetry and therefore $J_{z,r}(x,y)$ is anti-symmetric in $x$ and $y$, i.e., $J_{z,r}(x,y) = - J_{z,l}(-x,y)$ and $J_{z,r}(x,y) = - J_{z,l}(x,-y)$, see~\cite{Reddy_SHG_TO,Optimization}; As shown in~\cite{Optimization}, this yields a quadrupolar field pattern. For simplicity, from now on we can only relate to the right cylinder. 

\section{The analytical solution in the slab frame}
We now apply the TO procedure used previously in~\cite{Reddy_SHG_TO} to study SFG. Specifically, since $H_z^{\omega_3}$ is preserved under the conformal inversion transformation~\cite{Reddy_SHG_TO}, we can simplify the TWs problem by using the inversion conformal transformation to transform from the touching dimer frame to the slab frame, solve the resulting (simpler) equations, and transform back, see Fig.~\ref{TWandslab}. 

First, we employ the inversion conformal transformation, given by
\begin{eqnarray}
x = \frac{g^2u}{u^2+v^2},~ y = \frac{g^2v}{u^2+v^2},\\
u = \frac{g^2x}{x^2+y^2},~ v = \frac{g^2y}{x^2+y^2},
\end{eqnarray}
where $g^2$ is a scaling constant and $u$ and $v$ are the transformed frame coordinates.
Hence, after injecting Eq.~(\ref{EbCwithJ}), Eq.~(\ref{eq:JR}) and Eq.~(\ref{eq:JL}), the boundary conditions~(\ref{EbC}) take the form
\begin{multline} 
\left(\frac{v^2+d^2}{g^2}\right) \partial_u \left[\frac{H_{z,sl}^{\omega_3}|_{bg}}{\varepsilon_{bg}} - \frac{H_{z,sl}^{\omega_3}|_m}{ \varepsilon_m^{\omega_3}}\right] = \\ \pm i \omega_3 \varepsilon_0  \Delta_{r/l}(u=\pm d,v) e^{{-i\alpha_s|v|}/{d}},\quad u = \pm d,\label{eq:solrightslab}\\ 
\end{multline}
where $d$ is the distance from the interface to the origin, see Fig.~\ref{TWandslab}, and $\Delta_{r,l}(u,v)$ are defined as 
\begin{eqnarray}
\Delta_r(u,v) &\equiv& \mathcal{R}\left(\frac{g^2u}{u^2+v^2},\frac{g^2v}{u^2 + v^2}\right), \\
\Delta_l(u,v) &\equiv& \mathcal{L}\left(\frac{g^2 u}{u^2 + v^2},\frac{g^2v}{u^2 + v^2}\right).
\end{eqnarray}
According to the anti-symmetric relation of the magnetic surface currents, $\mathcal{R}(x,y)$ and $\mathcal{L}(x,y)$ are anti-symmetric as well, hence, the boundary conditions in Eq.~(\ref{eq:solrightslab}) are identical.

In order to calculate the fields on the boundaries, we adopt the slowly varying amplitude approximation and define an ansatz solution, similar to~\cite{Reddy_SHG_TO}, based on the following considerations:
\begin{itemize}
\item Since we expect surface plasmon waves to occur, the solution must have exponential decay (along the transverse coordinate $u$), while the longitudinal dependence (along $v$) has to be oscillatory.

\item The anti-symmetric source (the magnetic current density~(\ref{magneticcurrents})) dictates an anti-symmetric solution.

\item The propagation constant of the solution (i.e., the $y$ -dependence) has to be the same as that of the source $\alpha_s$ defined as
\begin{eqnarray}
\alpha_s = \frac{\alpha^{\omega_1} + \alpha^{\omega_2}}{2}.
\end{eqnarray}
\end{itemize}

Accordingly, the ansatz solution will be
\begin{eqnarray} \nonumber
H_{z,sl}^{\omega_3} &=& \frac{-i\omega_3\varepsilon_0  g^2d}{\alpha_s\mathcal{P}}\left(\frac{\Delta_r(u=d,v)}{ v^2+d^2}\right)~e^{i\alpha_s|v|/d} \times \\&&
\begin{cases}
\sinh\left(\alpha_su/d\right),\quad |u|< d,\\
\textrm{sgn}[u]\sinh(\alpha_s)~e^{\alpha_s\left(1-|u|/d\right)},\quad |u|>d,
\end{cases}
\end{eqnarray}   

where $\mathcal{P}$ is the so-called Phase Matching (PM) factor, given by~\cite{Reddy_SHG_TO} \begin{eqnarray*}
\mathcal{P} = \cosh\left(\alpha_s\right) + \frac{\varepsilon_{bg}}{\varepsilon^{\omega_3}_m} \sinh \left(\alpha_s\right).
\end{eqnarray*} .
\section{The analytical solution in the TW frame}
As mentioned above, $H_{z,sl}^{\omega_3}$ is preserved under inverse conformal transformation back to the TW frame, thus, we can now find the solution in the touching dimer simply by transforming back into that frame. This gives
\begin{eqnarray}
&&H_{z,TW}^{\omega_3} (x,y) = \frac{- 4i \omega_3 \varepsilon_0 a  }{\alpha_s \mathcal{P} }\left(\frac{ r^4\mathcal{R}\left(\tau_x,\tau_x\right)}{4 a^2 y^2 + r^4}\right)~e^{i{2a \alpha_s|y|}/{r^2}} \times \nonumber \\  && 
\begin{cases} 
\sinh\left({2a\alpha_sx}/{r^2}\right),\quad r^2 + 2a|x| > 0,\\
\textrm{sgn}[x] \sinh{{\alpha_s}}~e^{\alpha_s\left(1-{2a|x|}/{r^2}\right)}, \quad r^2+2a|x| < 0, \end{cases}\label{eq:nearfieldsol}
\end{eqnarray}
where $r^2 = x^2 + y^2$, such that $r^2 + 2 a |x| > 0\ (< 0)$ correspond to regions outside (inside) the TWs, and $\tau_x$ and $\tau_y$ are coordinates defined as 
\begin{multline}
(\tau_x,\tau_y) = \left( \frac{1/(2a)}{1/(4a^2) + y^2/(x^2 + y^2)^2}, \right. \\ \left.
\frac{y/(x^2 + y^2)}{1/(4a^2) + y^2/(x^2 + y^2)^2}  \right).  \label{eq:tau}
\end{multline}
These coordinates map points in the domain outside the TWs to the TWs perimeter~\cite{Optimization}. Since $H_{z,TW}^{\omega_3}$ is continuous in the entire space, it would be interesting to find the spatial variation of $H_{z,TW}^{\omega_3}$ on the perimeter of the cylinders close to the touching point. This arc obeys $x^2 + y^2 = 2ax$, so that
\begin{eqnarray}
H_{z,TW}^{\omega_3}(x,y) =- i \omega \varepsilon_0\left[\frac{\sinh \alpha_s}{\alpha_s\mathcal{P}}\right] ~x~\mathcal{R}(x,y)~ \exp\left( \frac{2ia\alpha_s y}{x^2 + y^2}\right).
\end{eqnarray}
Using Eq.~(\ref{eq:JR}), this equation can be rewritten as
\begin{eqnarray}
H_{z,TW}^{\omega_3}(x,y) =- i \omega \varepsilon_0\varepsilon_{bg}\left[\frac{\sinh{\alpha_s}}{\alpha_s \mathcal{P}}\right]~x~ J_{z,r}(x,y). \label{sh_sol_simp}
\end{eqnarray}
This is the generalization of the solution presented in~\cite{Reddy_SHG_TO} to the (nondegenerate) case of SFG.

As previously in~\cite{Reddy_SHG_TO}, we support the analytic result~(\ref{sh_sol_simp}) with numeric simulations based on finite element method using commercially available package COMSOL Multiphysics (V 3.5a), see discussion in~\cite[Appendix E]{Reddy_SHG_TO}. We again observe excellent agreement, see Fig.~\ref{TWandslab}(c).

\section{Discussion}

\begin{figure}[H]
\centering
\includegraphics[width=0.475\textwidth]{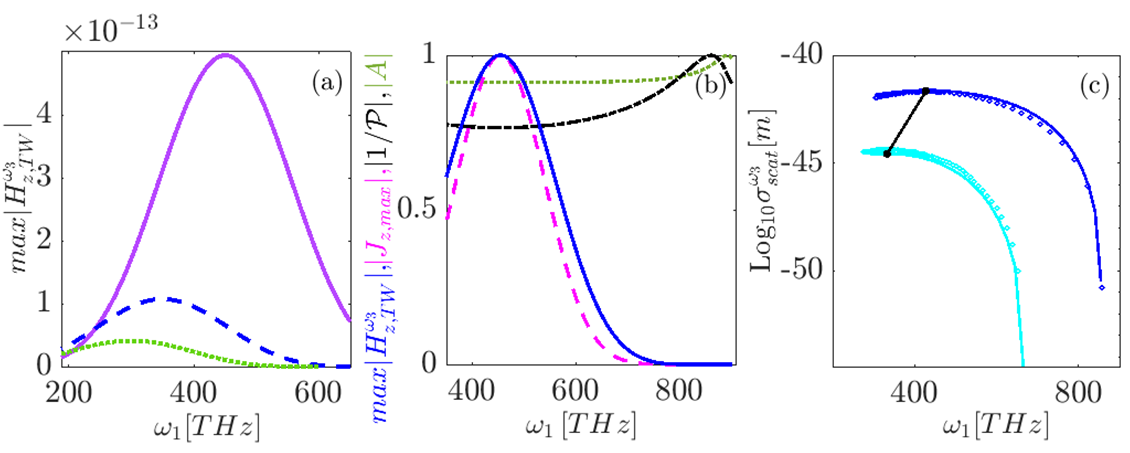}
\caption{(Color online) (a) The analytic solution of $max\left|H_z^{\omega_3}\right|$ for  $\omega_3 = 900$~THz (solid purple line), $700$~THz (dashed blue line), $600$~THz (dotted green line) The degenerate cases, $\omega_1 = 450,350,300$~THz, are marked by the dashed lines. (b) Comparison of the elements (normalized absolute value) of the analytic solution, Eq.~(\ref{sh_sol_simp}), at $\omega_3 = 909$ THz. $A$ (dotted green line) represents the the amplitude, $1/\mathcal{P}$ represents the phase matching element (dash-dotted black line), $max(H_z)$ is the maximum SFG magnetic field (solid blue line) and $max(J_z)$ (dashed magenta line) is the maximum magnetic current.(c) Analytic (line) and numeric (dots) scattering cross-section. Two plots for fixed SFG frequencies $\omega_3 = 850$ and $670$ THz, represented by the blue and cyan lines respectively. The black line represents the trajectory of the SHG combinations.} \label{SFGanalysis}
\end{figure}

Having obtained the solution for the SFG, we can now find the conditions for optimal efficiency. To do that, we first calculate the maximum magnetic (near-)field, $max\left|H_z^{\omega_3}\right|$, on the perimeter of the right cylinder for all the $\omega_{1,2}$ combinations in range of $200 - 900$ THz that yield a fixed value of $\omega_3$. We find that the strongest response is obtained in the degenerate case (SHG), i.e., when $\omega_1 = \omega_2$, see Fig.~\ref{SFGanalysis}(a). 
We can also go beyond the near-field calculation of~\cite{Reddy_SHG_TO} and investigate if the far-field calculations exhibit the same SHG superiority as the near-fields. To do that, using the recipe described in~\cite{Optimization}, we compute analytically and numerically the expressions for the scattering cross-section of the TWs for the SFG case, $\sigma^{\omega_3}_{scat}$. Specifically, we calculate the scattered power (on the contour $\mathcal{S}$ enclosing the TWs)
\begin{eqnarray}
P^{\omega_3}_{scat}~=~\frac12 \int_{\mathcal{S}} Re(E_{\phi}^{\omega_3}[H^{\omega_3}_z]^*) d\mathcal{S}, \label{Power}
\end{eqnarray}
where $E_\phi^{\omega_3}$ represents the parallel component of the  SFG electric field along the contour $\mathcal{S}$.
The SFG frequency was fixed for different values of $\omega_3 = 850$ and $670$ THz. In Fig.~\ref{SFGanalysis}(c), we observe an excellent agreement between the analytic and numeric solutions for the cross-sections. Furthermore, we also observe the superiority of the SHG in the far-field study. This result is consistent with the conclusion drawn in~\cite{Optimization} that $J_{z,r}$ is the dominant element determining the far-field response. 

This somewhat unexpected result is the main one of this work.

In order to understand why the SHG is superior over all other SFG combinations, we analyze the elements of the analytic solution~(\ref{sh_sol_simp}), see~Fig.~\ref{SFGanalysis}(b). On one hand, we observe that the amplitude $A = - i \omega \varepsilon_0 \varepsilon_{bg}\sinh{\alpha_s}/\alpha_s$ and the PM element are nearly independent of frequency, at least in the regime where the SFG peaks. On the other hand, we observed a similar trend for the $max\left|H_{z,TW}^{\omega_3}\right|$ and the maximum absolute magnetic current density, $max\left|J_{z,r}\right|$. Therefore, we conclude that $J_{z,r}$ has the dominant effect on $H_{z,TW}^{\omega_3}$.

\begin{figure}
\centering
\includegraphics[width=0.45\textwidth]{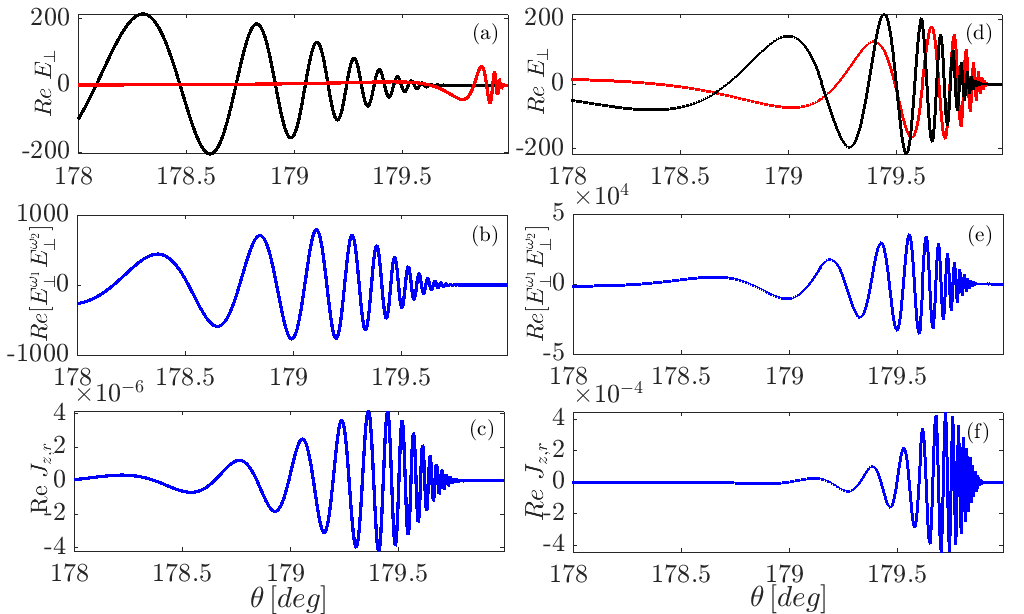}
\label{fig4:sub-second}
\caption{(Color online) (a) $Re\left[E_\perp\right]$ for $\omega_1 = 200$ THz (red solid line)and $\omega_2 = 708$ THz (black solid line), (b) $Re\left[E^{\omega_1}_\perp E^{\omega_2}_\perp\right]$ and (c) $Re\left[J_{z,r}\right]$.(d)-(f) Same as (a)-(c) for a smaller frequency difference, namely, $\omega_1 = 408$ THz (red solid line) and $\omega_2 = 500$ THz (black solid line).} \label{Jcompnentsanalysis}
\end{figure}

Thus, let us look more carefully at the constituents of the source term $J_{z,r}$. Eq.~(\ref{eq:JR}) shows that like in any 3 optical wave nonlinear interaction, the source is proportional to the product of the incident fields. The highly oscillatory nature of these fields close to the touching point means necessarily, that optimal overlap is obtained when the frequencies of the incident plane waves are identical. Otherwise, the overlap (and hence, the overall amplitude of the source) deteriorates with the frequency difference, see Fig.~\ref{Jcompnentsanalysis}. Thus, unlike the prior expectation that the additional degree of freedom enabled by SFG compared with SHG will yield better efficiency, we find that SHG is optimal for the near-fields of the TWs.

\section{Conclusion}
We have shown that the additional degree of freedom obtained by allowing the incoming frequencies to differ does not provide any additional improvement in the efficiency of the frequency conversion process. Our analysis shows that this originates from the highly oscillatory nature of the fields near the touching point. A similar conclusion is thus expected also for difference frequency generation~\cite{Che2016} or four wave mixing in such structures. Conversely, degenerate difference frequency generation (in particular, THz emission, see e.g.,~\cite{Fan_THz_MDM,Ellenbogen-Minerbi}) is thus expected to be efficient.

The analysis described in this work would be easily extended to the study of additional problems in nonlinear optics using TO, such as the study of non-local nonlinear effects (using e.g., the formulation of~\cite{Antonio_kissing_cyls_nonlocal_PRB,Antonio_kissing_cyls_nonlocal_PRL}), as well as the study of other structures such as non touching wires or other singular structures~\cite{Yu_singularities}, 3D particle configurations~\cite{3D_crescent} and (singular) gratings and 2D materials~\cite{Paloma-review-TO}.
\medskip

\noindent\textbf{Disclosures.} The authors declare no conflicts of interest.

\bibliography{REF}
\bibliographyfullrefs{REF}

\end{document}